\documentclass[twocolumn,3p]{elsarticle}

\usepackage{graphicx,amssymb,citesort}

\journal{Physica B}

\begin{document}
\begin{frontmatter}

\title{Bipolaron Density-Wave Driven By Antiferromagnetic Correlations and Frustration in Organic Superconductors\tnoteref{t1}}
\tnotetext[t1]{Supported by DOE Grant No. DE-FG02-06ER46315.}

\cortext[cor1]{Corresponding author}
\author[1]{R.T. Clay\corref{cor1}}
\ead{r.t.clay@msstate.edu}
\author[2]{H. Li}
\author[2]{S. Mazumdar}

\address[1]{Dept. of Physics \& Astronomy and HPC$^2$ Center for
Computational Sciences, Mississippi State University}
\address[2]{ Department of Physics, University of Arizona
Tucson, AZ 85721}

\date{\today}
\begin{abstract}
We describe the Paired Electron Crystal (PEC) which occurs in the
interacting frustrated two-dimensional $\frac{1}{4}$-filled band.  The
PEC is a charge-ordered state with nearest-neighbor spin singlets
separated by pairs of vacant sites, and can be thought of as a
bipolaron density wave. The PEC has been experimentally observed in
the insulating state proximate to superconductivity in the organic
charge-transfer solids.  Increased frustration drives a
PEC-to-superconductor transition in these systems.
\end{abstract}
\end{frontmatter}

Cuprate high $T_c$ and organic charge transfer solid (CTS)
superconductors share quasi-two-dimensionality (quasi-2D), strong
electron-electron (e-e) interactions and unconventional
superconductivity (SC).  Consequently, ideas first applied to the
cuprates, such as spin fluctuation-mediated SC, have also been applied
to the CTS \cite{Kontani08a}. In the present work we report exact
numerical calculations that cast severe doubt on this mechanism.

While antiferromagnetism (AFM) is adjacent to SC in some CTS, most
notably the $\kappa$-(ET)$_2$X, there are exceptions. In other CTS the
insulating phase adjacent to SC is charge ordered (CO) or has a spin
gap (SG), or both. In analogy to the spin fluctuation mechanism,
mean-field theories of charge fluctuation-mediated SC have been
proposed. Separate mechanisms for different classes of CTS
superconductors seem unlikely, given the similarities in crystal
structure and molecular makeup between CTS. In the second part of this
work we propose a single mechanism that can explain the exotic
insulating states and unconventional SC in the CTS.

\begin{figure}
\centerline{\resizebox{2.85in}{!}{\includegraphics{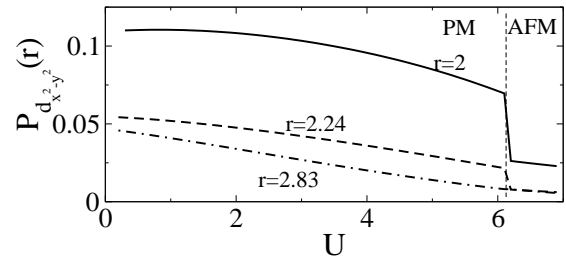}}}
\caption{Exact $d$-wave SC pair correlation functions
  versus distance $r$, for the $\frac{1}{2}$-filled $4\times$4 periodic
  anisotropic triangular lattice for $t^{\prime}/t=0.5$.  SC
  correlations similarly decrease monotonically with increasing $U$
  for all $t^\prime/t$ \cite{Clay08a}.}
\label{hubb12}
\end{figure}
SC in exotic superconductors often occurs at specific electron
concentrations, a feature that is beyond the scope of the BCS theory.
CTS superconductors have a carrier concentration $\rho$ of one hole or
electron per two molecules (i.e., $\rho$=0.5). We present here a
mechanism for frustration-driven transition from AFM to the PEC in the
$\rho$=0.5 2D CTS. Further increase in frustration drives a
PEC-to-superconductor transition that we believe gives the proper
mechanism of SC in the CTS. We believe that our mechanism of SC
applies to other exotic superconductors that share carrier density of
0.5, strong e-e and electron-phonon (e-p) interactions, and lattice
frustration with the CTS.

Mean-field calculations find a region of SC
between AFM and metallic phases in the $\frac{1}{2}$-filled
anisotropic triangular lattice Hubbard model. We report exact numerical
calculations that find no indication of SC. The Hamiltonian is:
\begin{equation}
H=-\sum_{i,j}t_{ij}(c_{i,\sigma}^{\dagger}c_{j,\sigma}+H.c.)
+U\sum_i n_{i,\uparrow}n_{i,\downarrow}.
\label{ham12}
\end{equation}
We consider a 2D 4$\times$4 square lattice with nearest-neighbor
(n.n.) hopping integral $t$ and next-nearest neighbor diagonal hopping
integral $t^\prime$, periodic in all directions.  In Fig.~\ref{hubb12}
we plot the d$_{x^2-y^2}$ pair-correlation function $P(r)$ as a
function of $U$ for several different distances $r$ measured in units
of the lattice constant.  The strength of the SC pair-correlation
functions decrease monotonically with increasing $U$. A weak increase
in onsite and n.n. $P(r)$ (not shown) simply reflects the increase in
strength of n.n. AFM correlations with $U$.  Other numerical
approaches have reached the same conclusion
\cite{Mizusaki06a,Tocchio09a}.  Claims of SC within the model appear
to be artifacts of the mean field approximation. We believe that
mechanisms of ``charge-fluctuation mediated SC'' based on mean-field
approximations also suffer from similar problems.

\begin{figure}
\centerline{\resizebox{2.0in}{!}{\includegraphics{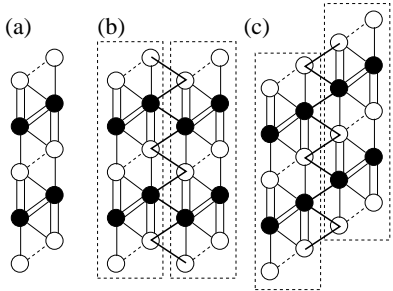}}}
\caption{(a) PEC in the $\rho=0.5$ zigzag ladder \cite{Clay05a}. Ladders can
be combined to form a 2D lattice in two ways, forming either
horizontal stripe (b) or diagonal stripe (c) CO patterns.
Black and white circles denote charge-rich and
charge-poor sites. Double bonds between the charge-rich sites indicate spin
singlets.}
\label{ladders}
\end{figure}
We now discuss $\rho=0.5$, which we believe is more appropriate to the
CTS.  For not too strong n.n. Coulomb interaction $V$, the ground
state in 1D in the presence of e-e and e-p interactions is CO with
charge pattern $\cdots$1100$\cdots$, where 1 (0) indicates charge
$0.5+\delta$ ($0.5-\delta$). Sites labeled `1' are coupled by a
singlet bond in the spin-gapped spin-Peierls (SP) state found quasi-1D
materials \cite{Clay03a}. This 1D bond-charge density wave is the
simplest example of the PEC. It is important to realize that the PEC
is driven by both e-p interactions and the AFM couplings.  A very
different kind of PEC forms when 1D chains are coupled together to
form a zigzag ladder, in which each site is bonded to a pair of
intrastack and a pair of interstack sites (see
Fig.~\ref{ladders}(a)). The ground state is again the same
$\cdots$1100$\cdots$ PEC, which however, now occurs along the zigzag
direction of the ladder (see Fig.~\ref{ladders}(a)) and the spin
singlet is {\it interchain}.  \cite{Clay05a}.

The structure of the PEC in the $\rho=0.5$ zigzag ladder gives us a
hint as to what could be expected in the triangular lattice at the
same carrier concentration.  In the $\frac{1}{2}$-filled $\rho=1$
case, rectangular ladders possess rung-based spin singlets. When
multiple ladders are coupled, however, the resultant 2D lattice is AFM
rather than SG, simply because each site now has 4 {\it singly
  occupied} n.n. sites \cite{Dagotto96a}.  The situation is even more
complex when $\rho=1$ zigzag ladders are coupled similarly to give a
coordination number 6. Very different scenarios emerge when $\rho=0.5$
zigzag ladders are similarly coupled, as is shown in
Figs.~\ref{ladders}(b) and (c).  Each ``occupied'' site continues to
have only 2 ``occupied'' neighbors, as in 1D. Thus stripe formation
(``horizontal'' or ``diagonal''), along with n.n. spin singlets are to
be {\it expected} in the 2D $\rho=0.5$ triangular lattice. In terms of
the n.n. $V$ interaction both CO patterns have the same static energy
\cite{Seo00a,Clay02a}. The strong tendency to spin-singlet formation
is a characteristic of the particular electron density as well
as of frustrated systems. Coupled rectangular ladders, for example,
have site populations of 0.5 each.

Based on the above, it is natural to speculate that the PEC persists
in the $\rho=0.5$ triangular lattice. We have recently confirmed this
within the 2D extended Hubbard Hamiltonian with n.n. e-e repulsion and
inter- and intrasite e-p couplings \cite{Li09a}.  We started with the
same periodic 4$\times$4 lattice as above, and performed
self-consistent calculations. For diagonal hopping $t^{\prime}$ less
than a critical $t^{\prime}_c$, the square lattice is dimerized along
one direction, with in-phase dimerization on all chains, and AFM
spin-spin couplings between the dimer unit cells. This explains the
AFM in the bulk of the $\kappa$-(ET)$_2$X. For $t^{\prime} >
t^{\prime}_c$ we find a clear frustration-driven transition from
the AFM to the PEC of Fig.~\ref{ladders}(b).  
We refer to the original work for details \cite{Li09a}.

\begin{figure}
\centerline{\resizebox{2.5in}{!}{\includegraphics{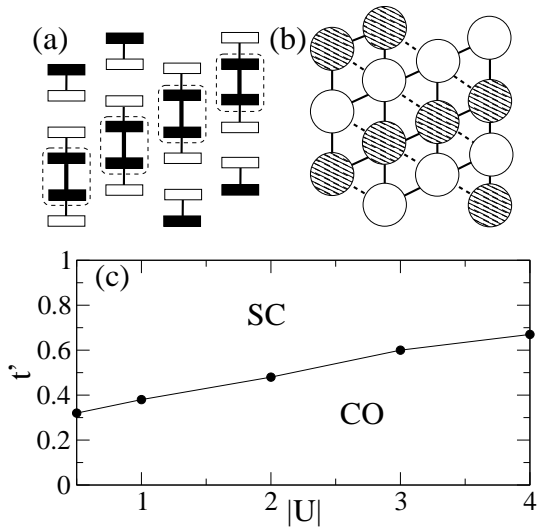}}}
\vskip 0.1in
\centerline{\resizebox{2.75in}{!}{\includegraphics{fig3c}}}
\caption{(a) PEC in the $\beta^\prime$ CTS. Singlets indicated by
  heavy line and dashed boxes.  (b) Effective $-|U|$, $+V$,
  $t^\prime$, $t$ model.  Pairs of singlet-bonded sites (vacancies)
  are mapped to a single doubly occupied (vacant) site in the
  effective model.  (c) Phase diagram of the effective model for $V=1$
  \cite{Mazumdar08a}.}
\label{negu}
\end{figure}

The PEC has been found in a number of 2D CTS with different
crystal structure, $\alpha$, $\beta^\prime$, $\theta$ and also $\kappa$.
\cite{Li09a}. The CO pattern is $\cdots$1100$\cdots$ along two
crystal directions and $\cdots$1010$\cdots$ along the third 
\cite{Li09a}. In the following we discuss the most notable cases.

The PEC shown in Fig.~\ref{ladders}(b) is found in the
$\theta$-(ET)$_2$MM$^{\prime}$(SCN)$_4$, where a SG occurs at T$_{SG}$
\cite{Mori06a}. The high T$_{SG}$ ($\sim$ 60 K) precludes a simple 1D
SP transition, which occurs at 10-20 K in the CTS. With decreasing
temperature the lattice parameter along the weakest hopping direction
decreases sharply, leading to increased frustration giving the
transition to the PEC \cite{Li09a}.

The PEC pattern in Fig.~\ref{negu}(a) is seen in some
$\beta^\prime$-X[Pd(dmit)$_2$]$_2$ \cite{Tamura09a}. Weakly frustrated
systems are AFM, but SGs occur in systems closer to being isotropic
triangular (X=Et$_2$Me$_2$Sb and X=EtMe$_3$P) \cite{Tamura09a}.  For
X=EtMe$_3$P, the charge densities and intermolecular distances are
exactly as in Fig.~\ref{negu}(a) \cite{Tamura09a,Li09a}.  Another CTS
with a similar structure, $\beta$-($meso$-DMET)$_2$PF$_6$, has the
same CO/bond arrangement \cite{Kimura06a}. Pressure-induced transition
to SC occurs in both systems.

Among the $\kappa$-(ET)$_2$X CTS, X=Cu$_2$(CN)$_3$ has a nearly
isotropic triangular lattice and does not display AFM order
\cite{Shimizu03a}. Although specific heat data appear to show no gap
\cite{Yamashita08a} thermal conductivity shows a small SG
\cite{Yamashita09a}. We propose that the ground state of this CTS is a
PEC, driven by frustration larger than that in the other
$\kappa$-salts. Because of near perpendicular orientations of
neighboring dimers, multiple ways of forming singlet-bonds between
charge-rich sites are possible, making any SG very small
\cite{Li09a}. Experimental evidences that support our proposal of CO
here include, (i) NMR line broadening at low temperature,
\cite{Kawamoto06a}, (ii) strong role of the lattice near the 6K
transition as seen in thermal expansion measurements \cite{Manna09a},
and (iii) frequency-dependent dielectric constant measurements that
indicate unequal charges within the ET dimers \cite{AbdelJawad09a}.
Note in particular that site charge occupancies 1=0...0=1 along the
``stacks'' provides a simple explanation of the
antiferroelectricity observed by these authors.

We have proposed that the $\rho=0.5$ PEC can be mapped to an effective
$\rho=1$ CO in which the singlet-bonded sites are replaced with an
effective doubly occupied single site (Figs.~\ref{negu}(a)--(b)) \cite{Mazumdar08a}.  The effective $\rho=1$ system is described
by a negative (attractive) $U$ and repulsive $V$. In agreement with
other authors, we assume that pressure increases frustration.  We have
shown that a CO-to-SC transition, driven by the frustration
$t^\prime/t$ occurs within this effective model (see 
Fig.~\ref{negu}(c)) \cite{Mazumdar08a}.

Other exotic $\rho$=0.5 superconductors, with proximate exotic
insulating states, include spinels LiTi$_2$O$_4$ \cite{Johnston73a},
and CuRh$_2$X$_4$ (X=S,Se) \cite{Hagino95a}. There are 0.5 d-electrons
per atom after Jahn-Teller distortion breaking the $t_{2g}$ orbital
symmetry.  There is evidence for PEC formation in CuIr$_2$S$_4$,
isostructural and isoelectronic with CuRh$_2$S$_4$; CO of
$\cdots$1100$\cdots$ form and n.n singlet formation are both found
\cite{Khomskii05a}.  Na$_{0.5}$CoO$_2$ is another example with both CO
and AFM phases at $\rho=0.5$ \cite{Foo04a}. Several common features,
viz., $\rho=0.5$, strong e-e interactions, lattice effects showing
electron-phonon (e-p) interactions, and geometrically frustrated
lattices, link these seemingly unrelated materials. Our proposed
theory unveils the relationship between them.

\end{document}